\documentclass[10pt,letterpaper]{article}
\usepackage[top=0.85in,left=2.75in,footskip=0.75in,marginparwidth=2in]{geometry}

\usepackage[utf8]{inputenc}

\usepackage{cite}
\usepackage{booktabs}

\usepackage{nameref,hyperref}

\usepackage{microtype}
\DisableLigatures[f]{encoding = *, family = * }

\raggedright
\usepackage{changepage}

\usepackage[aboveskip=1pt,labelfont=bf,labelsep=period,singlelinecheck=off]{caption}

\makeatletter
\renewcommand{\@biblabel}[1]{\quad#1.}
\makeatother

\usepackage{lastpage,fancyhdr,graphicx}
\usepackage{epstopdf}
\fancyheadoffset[L]{2.25in}
\fancyfootoffset[L]{2.25in}

\usepackage{color}

\definecolor{Gray}{gray}{.25}

\usepackage{graphicx}

\usepackage{sidecap}

\usepackage{wrapfig}
\usepackage[pscoord]{eso-pic}
\usepackage{amsmath}
\usepackage{amssymb}
\usepackage{mathtools}
\usepackage{upgreek}
\usepackage[fulladjust]{marginnote}
\reversemarginpar

\begin{document}
\vspace*{0.35in}

\begin{flushleft}
{\Large
\textbf\newline{Random sequential adsorption of spheres on a cylinder}
}
\newline
\\
Edvin Memet\textsuperscript{1, \dag},
Nabila Tanjeem\textsuperscript{2, \dag},
Charlie Greboval\textsuperscript{2,3},
Vinothan N. Manoharan\textsuperscript{1,2},
L. Mahadevan\textsuperscript{1,2,4,5,*},
\\
\bigskip
\bf{1} Department of Physics, Harvard University, Cambridge, MA 02138, USA
\\
\bf{2} Harvard John A. Paulson School of Engineering and Applied Sciences, Harvard University, Cambridge, MA 02138, USA
\\
\bf{3} Sorbonne Universit\'e, CNRS, Institut des NanoSciences de Paris, INSP, F-75005 Paris, France
\\
\bf{4} Department of Organismic and Evolutionary Biology, Harvard University, Cambridge, MA 02138, USA
\\
\bf{5} Kavli Institute for Nano-Bio Science and Technology, Harvard University, Cambridge, Massachusetts 02138, USA
\bf{\dag} These authors contributed equally
\\
\bigskip

* lmahadev@g.harvard.edu

\end{flushleft}

\section*{Abstract}
{Inspired by observations of beads packed on a thin string in such systems as sea-grapes and dental plaque, we study the random sequential adsorption of spheres on a cylinder. We determine the asymptotic fractional coverage of the cylinder as a function of the sole parameter in the problem, the ratio of the sphere radius to the cylinder radius (for a very long cylinder) using a combination of analysis and numerical simulations. Examining the asymptotic structures, we find weak chiral ordering on sufficiently small spatial scales.  Experiments involving colloidal microspheres that can attach irreversibly to a silica wire via electrostatic forces or DNA hybridization allow us to verify our predictions for the asymptotic coverage.}


\section*{Introduction}

Adsorption processes in which particles are randomly deposited on an extended substrate can occur in a broad range of physical, chemical, and biological systems, such as binding of ligands on polymer chains, chemisorption, physisorption, coating, paint, filtration, designing composites, drug delivery, and solid-state transformations \cite{evans1993,wang1996,osberg2015, shelke2008, adamczyk1991, tanemura1980}. On the one hand, the monolayer adsorption of small molecules is usually described by an equilibrium picture, resulting from adsorption-desorption kinetics, particle hopping, or diffusion \cite{evans1993,thompson1991}. On the other hand, larger molecules (proteins, viruses, bacteria, colloids, cells) may interact with the surface so strongly that they exhibit virtually no desorption, surface diffusion, or reaccomodation, and do not interact with subsequently adsorbed molecules except for steric exclusion effects \cite{talbot2000, rosen1986, swendsen1981, thompson1991}. The irreversible particle deposition that occurs in such nonequilibrium systems can be modeled as a random sequential adsorption (RSA) process \cite{thompson1991, bartelt1990}, also known as a ``car parking problem'' in the one-dimensional continuum case \cite{renyi1963}. 

Two natural questions are of central interest in RSA \cite{privman1991, ciesla2014, tarjus1991}. The first is the surface coverage fraction $\rho_{\infty}$ -- the ratio of surface covered by adsorbed particles to the total collector area in the longtime limit, when there is no more space for additional particles to adsorb.  This surface coverage fraction is smaller than the close-packing density. The second is the kinetics of particle adsorption $\rho \left(t\right)$. The answers to both these questions depend on the geometry, dimensionality, size, and shape of the particles being adsorbed \cite{shelke2008}. 

The simplest version of the problem is the case of uniform size segments being adsorbed on an infinite line, solved analytically by Renyi \cite{renyi1963}. Since then, many variants of this problem have been studied - by considering the role of dimensionality and shape of both the particles and the substrate as well as particle size distribution. These include considerations of heterogeneous 1D particles on 1D substrates, 1D particles on flat 2D substrates (needles on a plane\cite{viot1992, talbot2000}, polymer chains on a lattice\cite{becklehimer1994, wang1996}, dimers on a ladder \cite{fan1992, evans1992}), 2D particles on flat 2D substrates (disks, rectangles/ellipses\cite{brosilow1991, viot1992} with fixed or arbitrary orientation, stars and other concave objects\cite{ciesla2014b}, mixed concave/convex objects\cite{shelke2008}, or compound objects\cite{ciesla2014} on a plane or on a narrow strip\cite{suh2008}); 3D particles on fractals\cite{ciesla2012} or porous solids\cite{thompson1991}; and 3D particles on flat 2D substrate (polydisperse spheres on a plane \cite{danwanichakul2007}).  

Here we consider the adsorption of spheres on a cylindrical wire, inspired by a range of biological systems that exhibit such a morphology, such as those shown in Fig.~1. These include dental plaque which exhibits a "corncob" morphology (Fig.~1A), comprised of streptococci held by an extracellular polysaccharide matrix on large filamentous bacteria \cite{manton2013, lakshman2006}, and fruits such as peppercorn, winterberry, and seagrapes (Fig.~1C,D), where phyllotaxis may also be relevant \cite{mughal2017}. Despite the commonality of these observations, there seem to be few studies on the adsorption of objects onto curved substrates. 

Some exceptions include parking on spheres \cite{rosen1986, mansfield1996, adamczyk1991, schade2013, feder1980, sadowska2016, chen2017}, hyperboloids\cite{chen2017}, projective planes\cite{chen2017}, and cylinders \cite{adamczyk1991}. Previous studies \cite{adamczyk1991, sadowska2014} have argued that the asymptotic coverage $\rho^{\mathrm{cyl}}_{\infty}$ for random parking of spheres of radius $R$ on a cylinder of radius $r$ and length $L$ can be related to that of disks on a flat plane by effectively unrolling the cylinder that passes through the centers of the particles: $\rho^{\mathrm{cyl}}_{\infty} \coloneqq N_{\infty} \pi R^2 / \left(2 \pi r L\right) =  \rho^{\mathrm{plane}}_{\infty} \left(1 + R/r \right)$, where $N_{\infty}$ is the number of particles adsorbed in the infinite time limit. This approximation is valid for relatively small values of $R/r$ but breaks down as particles become larger compared to the cylinder ($R/r \rightarrow \infty$) -- in other words, when wire curvature becomes important. Here, we do not limit ourselves to the weak curvature regime and use a combination of analysis, simulations, and experiments, to characterize the asymptotic coverage of spheres on a rigid wire as a function of $R/r$. 




\begin{figure}[h!]
\centering{
\includegraphics[width= 0.8 \linewidth]{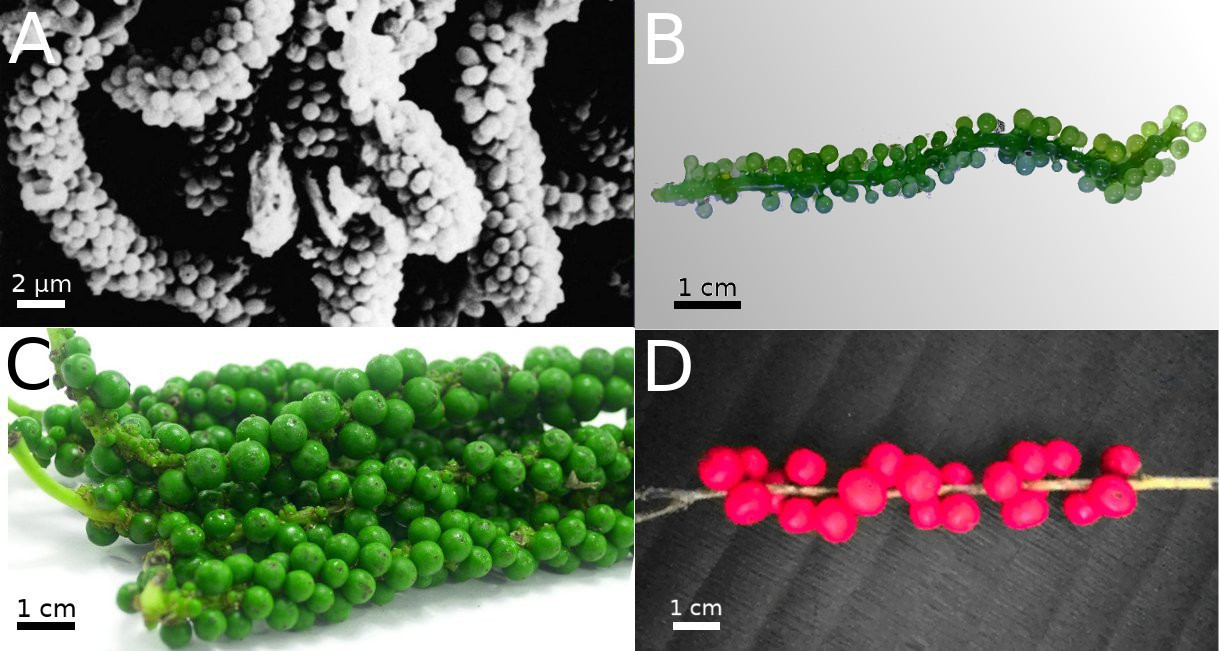} 
}
\caption{Spheres-on-cylinder morphologies in (A) dental plaque ``corncob'' formations\cite{lakshman2006}, (B) sea grapes, (C) peppercorn drupes, and (D) winterberries.}
\end{figure}

\section*{Results}

We start by noting that the adsorption of spheres of radius $R$ on a cylinder of radius $r$ is characterized by two degrees of freedom -- the axial coordinate $z$ and the azimuthal angle $\phi$ (Fig.~2A). Consequently, it is possible to map the 3D geometry of this process onto a 2D adsorption problem in the $\phi - z$ plane, where spheres are characterized by an angular envelope with an extent that depends on the axial coordinate. Such a transformation has been used previously in studies pertaining to the packing of spheres inside a cylinder \cite{pickett2000, mughal2012, chan2011}. To find this angular envelope, consider a sphere whose center is located at $ (z_0, \, \phi_0)$. In a horizontal slice at height $z$, the radius of the circle is $s(z-z_0)$  $= \sqrt{R^2 - (z-z_0)^2}$ and its angular extent is $\Delta \phi = \arcsin \left( s(z-z_0) / (R+r) \right)$ (Fig.~2A). This gives us an equation for $\phi (z)$ -- that is, the shape of a sphere in the $\phi - z$ plane:
\begin{equation}
  \begin{split}
    \phi (z-z_0) &= \phi_0 \pm \Delta \phi \\
    &= \phi_0 \pm \arcsin \left[\frac{\sqrt{R^2 - (z-z_0)^2}}{(R+r)}\right],
  \end{split}
\end{equation}
\noindent where $z-z_0 \in [-R, \, R]$. 

\begin{figure}[h!]
\centering{
\includegraphics[width= 0.95\linewidth]{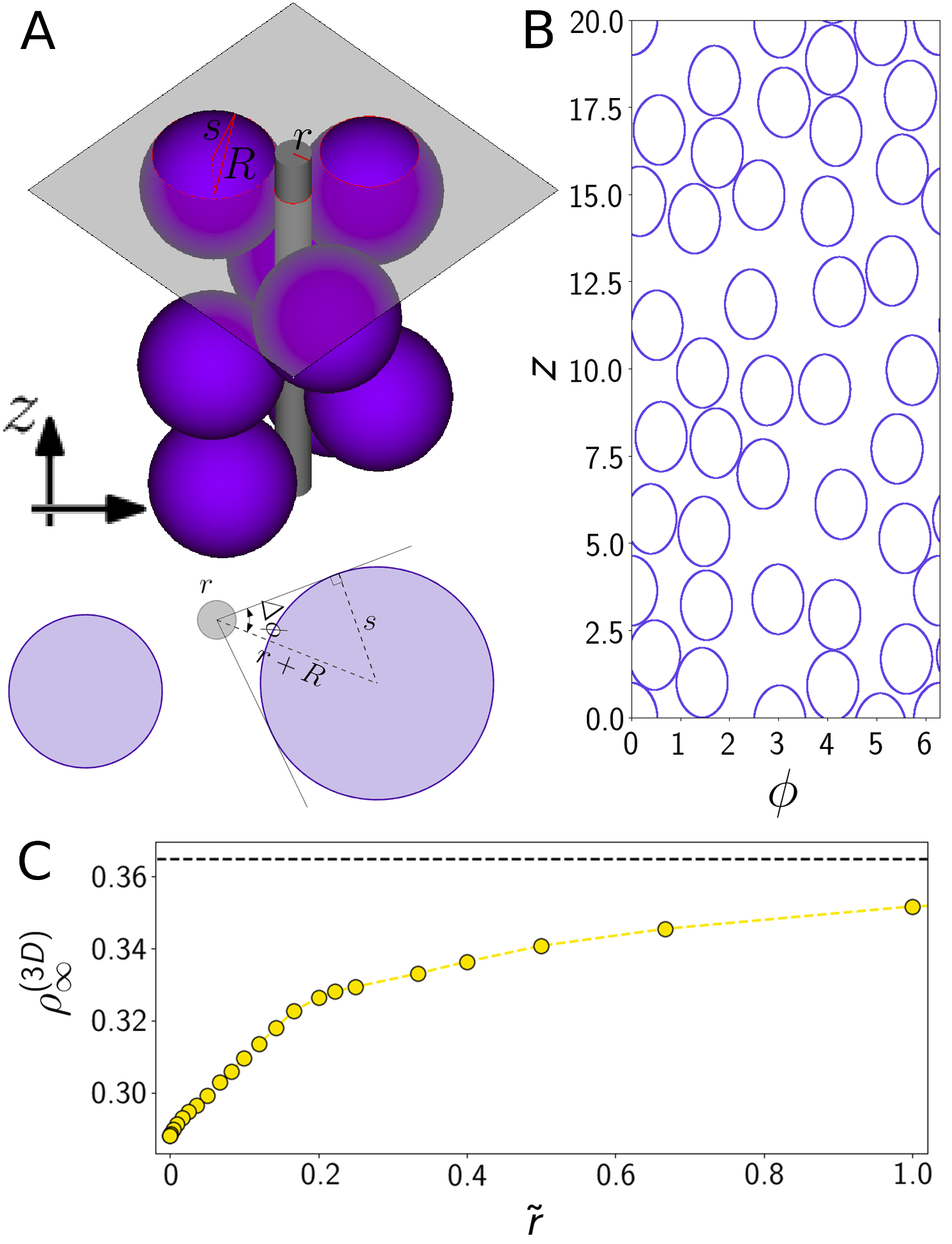} 
}
\caption{(A) Cartoon of spheres adsorbed on a wire. A sectioning plane indicated through the shade change is shown below, with $r$ indicating the cylinder radius and $R$ the particle radius, $\rho$ indicating the radius of a particular cross-section through the particle, and $\Delta \phi$ indicating the angle subtended by the particle cross-section at the center of the wire. (B) Two-dimensional representation in the $\phi - z$ plane of spheres of radius $R=1$ adsorbing on a cylinder of radius $r=1$ and length $L=20$. (C) Longtime coverage $\rho^{(3D)}_{\infty}$ versus scaled wire size $\tilde r = r / R$ (Eq.~(\ref{eq:rho3d})) from effective 2D simulations of spheres parking on a cylinder. The dashed black line indicates the longtime coverage for random sequential adsorption of spheres on a plane, $\lim_{\tilde  r \to \infty} \rho^{(3D)}_{\infty} = (2/3) \lim_{\tilde  r \to \infty} \rho^{(2D)}_{\infty}  \approx 0.3647$, where $ \lim_{\tilde  r \to \infty} \rho^{(2D)}_{\infty} \approx 0.5471$ is the asymptotic coverage of discs on a plane~\cite{ciesla2018}.}
\end{figure}

Consequently, random sequential adsorption of spheres on a cylinder is equivalent to random sequential adsorption of oblong 2D objects with shape given by Eq. (1) on a 2D strip of width $ 2 \pi$ and length $L$ (Fig. 2B). We simulate the latter process for different values of the ratio between cylinder and particle size $\tilde r$ $ \equiv r /R$, using periodic boundary conditions. For each deposition attempt we generate a pair of random numbers $(\phi, \, z)$ representing a potential adsorption site and check for overlaps with existing particles in the vicinity. In the absence of overlaps, the trial particle is successfully adsorbed and remains fixed thereafter; otherwise it is removed from the system. Since reaching a completely blocked state in which there remains no space for new particles to adsorb may take very long, we instead extrapolate $N_{\infty}$, the asymptotic number of particles deposited on the cylinder, from the longtime kinetics \cite{hinrichsen1986, ciesla2014b, brosilow1991, meakin1992}. That is, for particles with two degrees of freedom, $\rho_{\mathrm{\infty}} - \rho \left(\tau\right) \sim \tau^{-1/2},$ as $ \tau \rightarrow \infty$ \cite{baule2017},  where scaled time $\tau$ is the attempt number times the area fraction of a particle scaled by the total area of the substrate. In addition, to minimize boundary effects, we use a cylinder size $L$ that is large compared to the particle size $R$ \cite{ciesla2014}.  

To find the longtime coverage from $N_{\infty}$, we define coverage as the ratio of occupied to available area in the 2D $\phi - z$ space  (where ``area'' has dimensions of length, since it is defined in the $\phi - z$ plane). The area occupied by a sphere is $A = \tilde A \, R $ where $\tilde A $ is the dimensionless area, $\tilde A = 2 \displaystyle \int_{-1}^{1} \arcsin \left[ \frac{\sqrt{1-\tilde z^2}}{1+\tilde r} \right] \mathrm{d} \tilde z$, expressed in terms of the scaled axial coordinate $\tilde z = \left(z-z_0\right)/R$ and scaled cylinder radius $\tilde r = r / R$. If we let $N$ be the total number of spheres adsorbed up to the current time and $L\rightarrow \infty$ be the length of the cylinder, the 2D coverage density is given by 
\begin{equation}
  \rho^{(2D)} \left( \tilde r\right) = \frac{N \tilde A R}{2 \pi L} = \frac{1}{2 \pi} \frac{N}{\tilde L} \tilde A = \frac{ \lambda \left(\tilde r\right) \tilde A \left(\tilde r\right)}{2 \pi},
  \label{eq:rho2d}
\end{equation}
where $\tilde L = L/R$ is the scaled cylinder length and $\lambda = N / \tilde L$ is the scaled linear particle density. We note that in the limit $\tilde r \rightarrow 0$, we have $\tilde A \rightarrow 4$ and $\rho^{(2D)} \rightarrow 2 \lambda \left(\tilde r \right) /\pi$, while as $\tilde r \rightarrow \infty$, $\tilde A \rightarrow \pi / \tilde r$ and $\rho^{(2D)} \rightarrow \lambda \left(\tilde r \right) / \left(2 \tilde r\right)$. 


To help facilitate the comparison to experiments, we convert the surface coverage $\rho^{(2D)}$ to a three-dimensional coverage representing the ratio between the total volume of the adsorbed spheres and the total available volume -- that is, the volume of the cylindrical annulus of inner radius $r$ and outer radius $r+2R$:
\begin{equation}
  \rho^{(3D)} \left( \tilde r\right) = \frac{4/3 N \pi R^3}{4\pi R L \left(  R +  r \right)} = \frac{1}{3} \frac{N}{\tilde L} \frac{1}{1+\tilde r} = \frac{1}{3} \frac{\lambda \left(\tilde r\right)}{1 + \tilde r}.
  \label{eq:rho3d}
\end{equation}
We note that in the low-curvature regime, $\tilde r \rightarrow \infty$, $\rho^{(3D)} \rightarrow \lambda \left( \tilde r \right) / \left(3 \tilde r \right) \approx 0.3647 $ (Fig. 2C), while in the high-curvature regime, $\tilde r \rightarrow 0$, $\rho^{(3D)} \rightarrow \lambda \left( \tilde r\right) / 3$.

How do these densities compare to the maximum densities achieved by close-packed spheres on the surface of a cylinder? To get the maximum packing density as a function of the ratio of wire and particle size, we refer to the literature on packing spheres \textit{inside} cylinders \cite{mughal2012, fu2016, mughal2011}; as long as there is a single layers of particles inside the cylinder (that is, all the particles are in contact with the cylinder), the packing densities can be easily mapped to densities corresponding to packing \textit{on} the surface of a cylinder. Using this method we find that the ratio of random to close-packed densities varies around 0.62. For example, when $\tilde r \rightarrow 0$, the random parking density is around 0.288 (Fig. 2C), while the densest packing of spheres inside a cylinder twice as large as the particles is around 0.47, giving a ratio of 0.61. Similarly, when $\tilde r = 0.5$, the corresponding ratio is approximately 0.63.  

\begin{figure}[h!]
\centering{
\includegraphics[width= 0.78\linewidth]{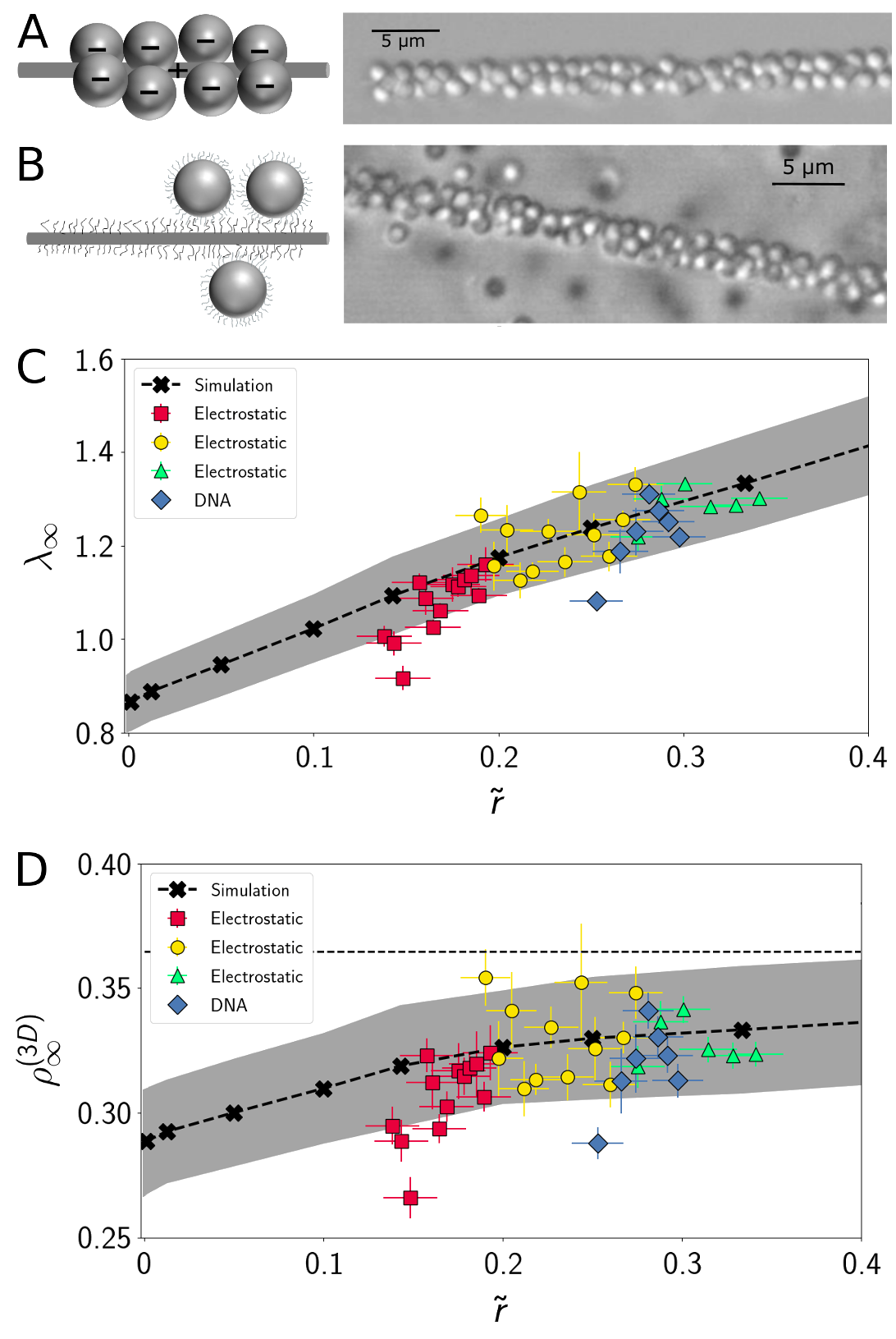} 
}
\caption{(A) Schematic (left) and optical micrograph (right) of negatively charged particles binding irreversibly to a positively charged nanowire (B) Schematic (left) and optical micrograph (right) of DNA-coated particles binding to nanowire coated with complementary DNA strands. (C) Linear particle density $\lambda$ versus scaled wire size $\tilde r = r/R$  and (D) 3D asymptotic density $\rho^{(3D)}_{\infty}$ versus scaled wire size $\tilde r = r/R$ from simulations (black crosses), and experiments using either electrostatic attraction (red squares, yellow circles, and green triangles show results from three different experiment samples) or DNA hybridization (blue diamonds). The dashed black line indicates, as in Fig. 2C, the longtime coverage for random sequential adsorption of spheres on a plane, $\lim_{\tilde  r \to \infty} \rho^{(3D)}_{\infty}   \approx 0.3647$. Shaded regions indicate an interval of two standard deviations from the mean simulation results. In order to reproduce the experimental uncertainty associated with small particles numbers, simulations were performed on short wire segments that accommodate around 50 particles.}
\end{figure}

To test our predictions, we designed an experimental system consisting of colloidal microspheres that can irreversibly attach to a wire (see Appendix for details of the protocols). An aqueous dispersion of microspheres, which have an average radius $R =  0.65\,\upmu \mathrm{m}$, is allowed to adhere to a silica wire via two different kinds of attractive interactions to drive the microsphere to adsorb: electrostatic and DNA-mediated interactions.    For electrostatic-mediated adsorption, we use oppositely charged microspheres and nanowires. The surfaces of the colloidal polystyrene microspheres contain negatively charged sulfate groups. To impart positive charge to the nanowire, we coat it with a cationic polyelectrolyte, poly-diallyldimethylammonium chloride (details in Appendix). After the colloidal microspheres are dispersed in water in the presence of the nanowire, they adsorb on the wire surface (Fig.~3A). 

For DNA-mediated adsorption, we functionalize the colloidal microspheres and silica nanowire with complementary DNA strands, which at room temperature can form strong, irreversible bonds between the wire and the particles (Fig.~3B). The DNA strands contain a dibenzocyclooctyne (DBCO) group, which binds to an azide group of a polymer layer deposited on the surfaces of the wire and the microspheres (details in Appendix). For both electrostatic and DNA-mediated adsorption, we do not observe particles diffusing on the surface or desorbing within the experimental timescale.  Furthermore, in both systems, the range of the interaction between adsorbed particles is much smaller than the particle size.  Thus, both experimental systems provide a reasonable realization of the random sequential adsorption process.  Using experiments on wires with different radii that range from 0.1--0.6 $\upmu m$ allows us to examine adsorption over a range of $\tilde r$ values. 

To find $\rho^{(3D)}_{\infty}$ experimentally, we count the number of particles $N_{\infty}$ adsorbed on a wire segment of length $L$ and average radius $r$ and compute the normalized linear particle density for that segment, $\lambda_{\infty} \left(\tilde r\right) =  N_{\infty} / \tilde L$. The results from experiments using electrostatic interactions are consistent between different samples (red squares, yellow circles, and green triangles in Fig.~3C) and with results obtained using DNA-mediated attraction (blue diamonds in Fig.~3C).

We see that our experimental results are consistent with results of simulations (Fig.~3C, black line) on wire segments of comparable length to the wire segments in the experiments (Fig.~3A and 3B). While a few experimental points lie outside the two-sigma interval (Fig.~3C, black shaded region) of the simulation results, these deviations are likely associated with the effects of energetics and kinetics.  In our analysis and simulations, all contacts are assumed to be point contacts, while in reality, the contact interactions have a non-zero range.  As a result, the energy of a contact decreases as the curvature of the wire increases. Consequently, the regime in which wires are highly curved (small $\tilde r$) is difficult to probe experimentally because weakly bound particles on highly curved wires can detach more quickly, leading to undersaturation. Indeed, the data are consistent with this -- the measured densities at the smallest $\tilde r$ values for both the electrostatic and DNA-mediated interactions fall below the simulated values (Fig.~3C and 3D, red squares and blue diamonds). Nonetheless, outside of these values, the experimental results agree with those of simulation, and, importantly, both the simulations and experiments show that the longtime coverage lies below that for random sequential adsorption on a plane (Fig~3D). These results validate our understanding of random sequential adsorption at weak to moderate wire curvature ($\tilde r \gtrsim 0.2$).

Although energetic and kinetic limitations prevent us from experimentally exploring the limit of high wire curvature ($\tilde r \lesssim 0.2$), we can understand this regime theoretically in terms of the effective parking of 2D shapes on the unwrapped cylinder (Eq.~(\ref{eq:rho2d})). Simulations in this regime reveal a surprising effect: $\rho^{(2D)}_{\infty}$ varies non-monotonically with wire radius (Fig.~4), in contrast to $\lambda_{\infty}$ and $\rho^{(3D)}_{\infty}$, both of which vary monotonically with $\tilde r$ (Figs.~2C, 3C, 3D).

 To understand this, we note that from  Eq.~(\ref{eq:rho2d}) we see that $\rho^{(2D)}_{\infty}$ is the product of two terms that depend on the scaled cylinder radius $\tilde r$. While $\tilde A = 2 \int_{-1}^{1} \arcsin \left[ \sqrt{1-\tilde z^2}/(1+\tilde r) \right] \mathrm{d} \tilde z$ necessarily decreases as $\tilde r$ increases, the asymptotic particle density $\tilde \lambda_{\infty} $ increases with $\tilde r$, since the maximum angular extent $\Delta \phi_{\mathrm{max}} = \sin^{-1} \left(1 / (1+\tilde r)\right)$ decreases, allowing for more spheres to fit around the wire. The contrasting behavior of the two terms indeed allows for the observed non-monotonicity in $\rho^{(2D)}_{\infty}$, though it does not guarantee it. In contrast, we find $\rho^{(3D)}_{\infty}$  increases monotonically with $\tilde r$ in simulations (Fig.~2C), even though Eq.~(\ref{eq:rho3d}), which expresses $\rho^{(3D)}_{\infty}$ as the ratio of two terms that both increase with $\tilde r$, does not exclude non-monotonic behavior.

Having examined the longtime coverage as a function of wire curvature, we now turn to the second quantity of interest, the kinetics of particle adsorption.  Interestingly, in idealized conditions where the energetics of the substrate-particle interaction play no role, there is a near-universal law (for convex particles) associated with the kinetic approach to the asymptotic coverage, described by a power-law form  \cite{shelke2008, vigil1990}
\begin{equation}
\rho_{\mathrm{\infty}} - \rho \left(\tau\right) \sim \tau^{-1/d_f},  \,\, \tau \rightarrow \infty,
\end{equation}
\noindent where time $\tau$ is defined so that each trial (adsorption attempt) corresponds to a scaled time increment equal to the area of an adsorbing particle scaled by the system area and $d_f$ represents the number of degrees of freedom of the object. In our system $d_f = 2$, meaning that we expect to find  $\rho_{\mathrm{\infty}} - \rho \left(\tau\right) \sim \tau^{-1/2}$. However, we also note that quasi-one-dimensionality emerges with increasing wire curvature, in that the 2D strip in $\phi - z$ space becomes increasingly narrow in the $\phi$ direction, which could change the asymptotic kinetics accordingly \cite{suh2008}. For example, in the limit $\tilde r \rightarrow 0$, the strip allows no more than two spheres to fit (tightly) around the cylinder at the same value of $z$. However, we do not observe a crossover to 1D asymptotic kinetics in our simulations as $\tilde r \to 0$, which suggests that the asymptotic exponent $d_f = 2$, independent of the scaled cylinder size $\tilde r$. 

\begin{figure}[h!]
\centering{
\includegraphics[width= \linewidth]{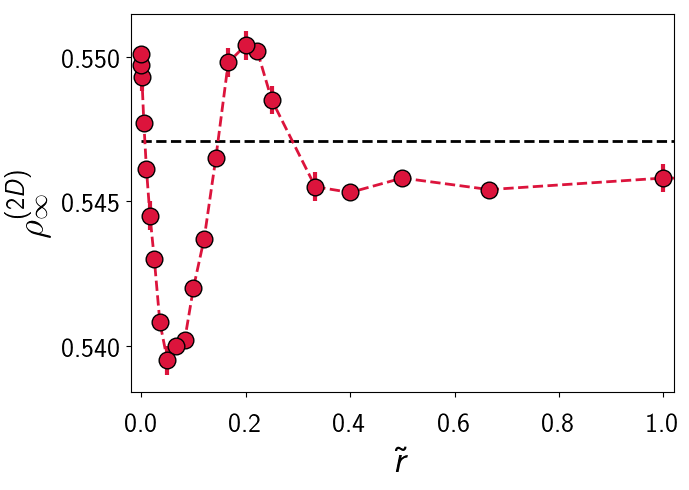} 
}
\caption{Longtime (asymptotic) coverage $\rho^{(2D)}_{\infty}$ versus scaled wire size $\tilde r = r / R$ (Eq.~(\ref{eq:rho3d})) from simulation. The dashed black line at 0.5471 indicates the longtime coverage for the random adsorption of discs on the plane (that is, in the limit $\tilde r \rightarrow \infty$).}
\end{figure}



\section*{Discussion}

Strongly and weakly charged colloidal chiral orderings in cylinders have been studied both theoretically and experimentally \cite{oguz2011, fu2017}. An interesting question is whether any chiral order emerges on short scales in our random adsorption process, where the system is not allowed to relax to its ground state. Long-time adsorption structures such as those shown in Fig. 5A suggest that there may be preferential alignment along certain directions. To quantify the alignment, we define an angular density-density correlation $\langle n \left(\vec r\right) \, n \left(\vec r', \theta \right) \rangle$ which counts pairs of particles whose centers are aligned at an angle $\theta$ within some tolerance $\Delta y$ (Fig. 5A, red shaded regions). An example of this correlation function is shown in Fig. 5B for the case in which we consider only pairs of particles within a range of $10 R$ from each other. From visual inspection of Fig. 5A, the density-density correlation function exhibits two moderate peaks at values of $\pm \theta_{max}$ symmetric around $\theta =0$. Sufficiently increasing the range over which we compute density correlations will eventually lead to the disappearance of the peaks. This is what we expect intuitively: as noted previously, random parking densities are only about 60 \% of the maximum close-packing densities, which suggests that ordering cannot survive on arbitrarily large scales. 

Repeating this analysis for different values of $\tilde r = r/R$, we find that the maximal density direction $\theta_{max}$ varies with $\tilde r$ (Fig. 5C, black crosses) such that the relationship between $\tan \theta_{max}$ and $\tilde r$ is linear: $\tan \theta_{max} \approx 3 \tilde r$ (Fig. 5C, red dashed line).  Very large particles ($R \gg r$ or $\tilde r \rightarrow 0$) tend to be slightly more aligned axially, while smaller particles (increasing $\tilde r$), with $\theta_{max} \rightarrow \pi/2$, tend to be more aligned radially (to take advantage of the additional space available due to substrate curvature).

In this letter, we have examined the random adsorption of large spherical particles on a thin cylindrical wire. By reducing the 3D adsorption problem to an effective 2D problem, we showed that curvature effects can indeed become significant and thus cannot be treated as a small perturbation. In order to test our predictions for the longtime particle density $\lambda_{\infty}$ as a function of the ratio $\tilde r$ of particle to wire size, we performed experiments in which the particle-wire interaction was mediated by electrostatics or DNA hybridization and showed the results are consistent with each other and with the results of simulations. While the high curvature limit was not accessible experimentally because of kinetic effects, simulations in this regime reveal an intriguing non-monotonic behavior of the 2D asymptotic coverage $\rho^{(2D)}_{\infty}$. Meanwhile, the 3D asymptotic coverage $\rho^{(3D)}_{\infty}$ varies monotonically with the ratio $\tilde r$ of wire to particle size, providing a recipe for designing structures with well-defined volume coverage simply by tuning the ratio of the radii. Accounting for the energetics of substrate and particle deformation is a natural, if complicated, next step. 

\begin{figure}[t]
\centering{
\includegraphics[width= 0.7 \linewidth]{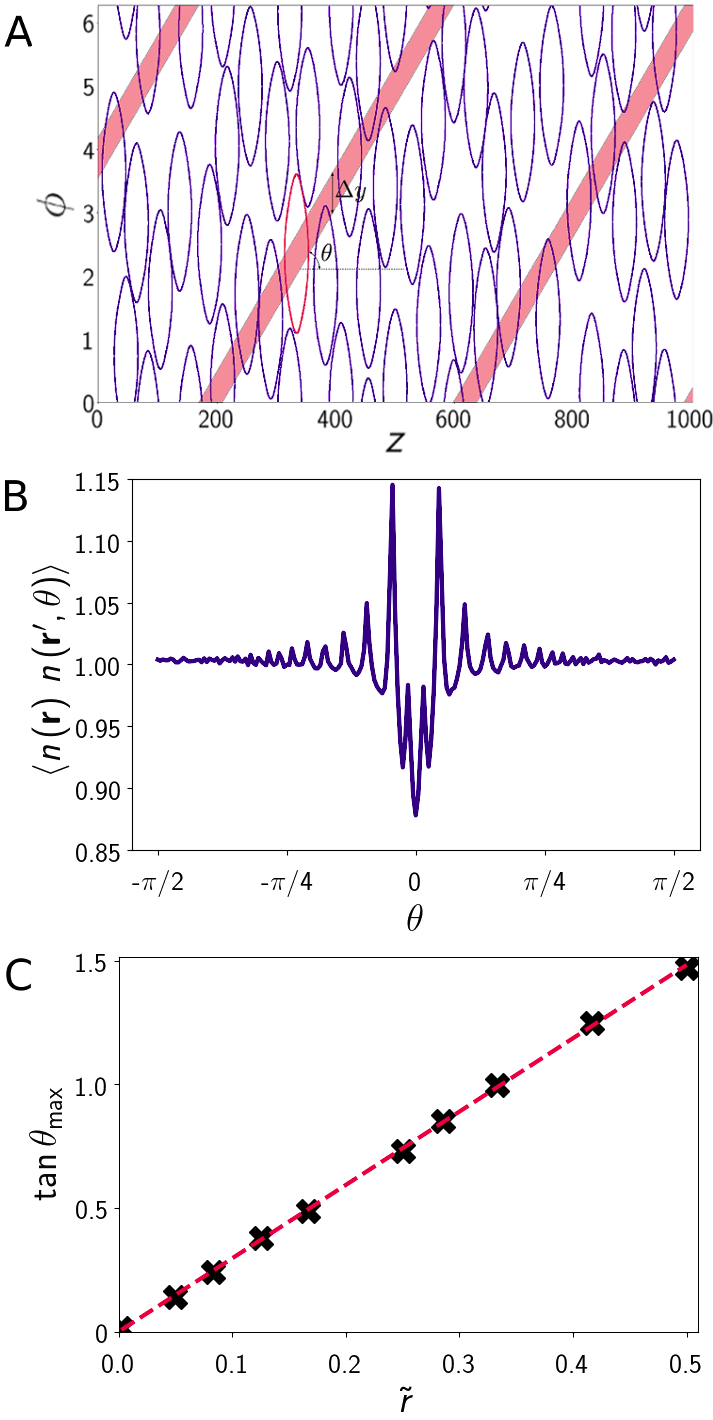} 
}
\caption{ (A) Two-dimensional representation in the $\phi - z$ plane of spheres of radius $R = 20$ adsorbing on a cylinder of radius $r = 1$ and length $L = 1000$. To compute the angular density-density correlation function, we fix a particle (shown in red), consider a strip of width $\Delta y$ along a direction $\theta$, and count the number of particles whose centers are contained inside this strip. Repeating this procedure for all particles and then for different angles $\theta$ yields (B) a plot of the angular density-density correlation $\langle n \left(\vec r\right) \, n \left(\vec r', \theta \right) \rangle$ as a function of angle $\theta$. (C) Varying $\tilde r = r/R$ changes the maximal density direction $\theta_max$ (that is, the angle for which  $\langle n \left(\vec r\right) \, n \left(\vec r', \theta \right) \rangle$ is maximum) as $\tan \theta_{max} \approx 3 \tilde r$.  
}
\end{figure}

 
\section*{Methods and Materials}

\subsection*{Nanowire fabrication}
We fabricate a thin cylindrical wire by tapering an optical fiber made of silica (supplied by Newport Corporation, part number F-SMF-28). First, we strip the outer layer from a piece of a fiber and clean the cladding by wiping it with isopropanol. We then attach the two ends of the fiber to two motorized stages and place a burner at the center of the fiber. Pulling the fiber by the motorized stages while the burner applies heat thins the wire down gradually until it eventually breaks\cite{tong2003,tong2005,tong2008}. The resulting wires are imaged with scanning electron microscopy, which allows us to measure the local wire diameter. Data from four different nanowires show that the change in wire diameter is approximately linear and gradual along the length of the fiber. Diameter variations for four different wires per 10$\,\mathrm{\upmu m}$ length are 1.76\,nm, 3.63\,nm, 5.12\,nm, and 3.85\,nm. 

\subsection*{Preparation of positively charged nanowire}
  We submerge the nanowire overnight in a 1\,M KOH solution to impart a negative charge to its surface. After the KOH treatment, we wash the nanowire five times with MilliQ water and transfer it to a solution of poly-diallyldimethylammonium chloride (shortened as polyDADMAC, purchased from Polysciences Inc., Molecular weight 240,000) in 20\,mM Tris-HCl buffer, prepared by mixing 28\% w/w polyDADMAC in water with 40\,mM Tris-HCl in 50:50 ratio and vortexing for 30\,s. After waiting 4--5 hours to allow the polyDADMAC to coat the nanowire, providing a positively charged surface, we take the nanowire out and wash it well with MilliQ water.
 
\subsection*{Preparation of negatively charged colloidal particles}
We purchase 8\% w/v sulfate-modified polystyrene particles (supplied by Molecular Probes, Life Technologies Inc.) with an average diameter of 1.3\,$\upmu$m and wash them three times by centrifuging at 4000\,$g$ and re-dispersing in MilliQ water. After the final wash, we disperse them in 0.05\,mM NaCl in water, resulting in a final particle concentration of 2\% w/v. 
 
\subsection*{Preparation of DNA functionalized nanowire}
To functionalize the silica nanowire with DNA oligonucleotides, we first clean it by overnight submersion in 1\,M KOH, then rinse with MilliQ water five times, and transfer it into (3-aminopropyl) triethoxysilane (APTES) solution. We prepare the solution by mixing 100\,mL methanol (99.9\%, supplied by VWR), 5\,mL glacial acetic acid (99.8\%, supplied by Acros Organics), and 3\,mL APTES (99\%, supplied by Sigma-Aldrich). After treating the nanowire in this solution for 30\,min, we rinse it with methanol and MilliQ water and transfer it to a PEG solution. The PEG solution is prepared by mixing NHS-PEG (5000 Da, supplied by Nanocs) and NHS-PEG-N$_3$ (5000 Da, supplied by Nanocs) in 10:1 ratio and dissolving them in 0.1\,M sodium bicarbonate buffer. We place the nanowire along with 192\,$\mathrm{\upmu L}$ of PEG solution between two glass coverslips and leave it overnight, at room temperature, so that the amino groups from APTES can form covalent linkages through $N$-hydroxysuccinimide (NHS) chemistry and form a PEG layer. The following day, we take the nanowire out from the PEG solution and rinse it with MilliQ water. Afterward, we attach DNA oligonucleotides to the NHS-PEG-$\mathrm{N_3}$ molecules on the nanowire surface by copper-free click chemistry \cite{chandradoss2014}. The DNA strands are 64-bases long and are synthesized with a dibenzocyclooctyne (DBCO) group on the 5'- end (purchased from Integrated DNA Technologies, HPLC purified). We put the nanowire in 10\,$\mathrm{\upmu M}$ of 168\,$\mathrm{\upmu L}$  DBCO-DNA (5'-T$_{50}$-AAGAGTAGGTTGATG-3') in phosphate buffer, sandwich it between two coverslips, and leave it for 24\,h before finally rinsing the nanowire with MilliQ water.

\subsection*{Preparation of DNA functionalized colloidal particles}
To coat the polystyrene microspheres with high density DNA brushes (5'-T$_{50}$-CCACATCAACCTACT-3') we incorporate a diblock copolymer made from polystyrene and poly(ethylene oxide)-azide into the microspheres and then attach DBCO-DNA to the azide functional group using copper-free click chemistry\cite{oh2015}.

\subsection*{Sample setup, imaging, and data anaysis}
After either functionalization method, we place the functionalized nanowire between two glass coverslips to make a sandwich sample chamber whose thickness is set to $67\,\mathrm{\upmu m}$ using mylar-film spacers. We then inject the suspension of particles and wait 10\,min before washing out excess particles. We wash by injecting a control solution with the same salt concentration as the colloidal suspension. We use NaCl for electrostatic screening in both cases: 0.05\,mM NaCl for electrostatic interactions and 200\,mM NaCl for DNA-mediated interactions.

Finally, we image the sample using a 60$\times$ water immersion objective.  Using optical microscopy images (Fig.~3A, 3B), we count the number of particles on wire segments of about 30\, $\mathrm{\upmu m}$ (chosen such that the segment diameters do not vary significantly along the length) and calculate  $\tilde N$. We assign error bars on $\tilde r = r/R$ based on the known polydispersity in particle size and the uncertainty in estimating the diameter of the wire. To calculate this uncertainty, we use the uncertainty in fitting and the difference in wire diameter between the two ends of the segment analyzed. Because the experimental images analyzed do not have periodic boundary conditions at the end of the wire segments, we find a number of particles that are only partially in the field of view of each analyzed segment. For each segment, we assign a lower limit of $\tilde N$ by not counting those particles and an upper limit by counting them. Then we calculate the mean and standard deviation of $\tilde N$ to assign an error bar to $\tilde N$.

\clearpage

\section*{Acknowledgments}

This work was supported by the Harvard MRSEC under National Science Foundation grant no.\@ DMR-1420570. We thank Dr. Gi-Ra Yi and Dr. Cheng Zeng for helpful discussions on developing experiment protocols.



\end{document}